\documentclass[sigconf]{acmart}

\AtBeginDocument{%
  }
\setcopyright{acmlicensed}
\usepackage{times} 
\usepackage{latexsym}
\usepackage[T1]{fontenc}
\usepackage[utf8]{inputenc}
\usepackage{microtype}

\usepackage{booktabs}
\usepackage{rotating}
\usepackage{multirow}
\usepackage[table]{xcolor}
\usepackage{pgfmath}
\usepackage{xfp}
\usepackage{makecell}

\usepackage[most]{tcolorbox}
\tcbuselibrary{listings,skins}
\usepackage{xcolor}
\usepackage{pifont}

\usepackage{pifont}

\usepackage{algorithmic}
\usepackage{textcomp}
\usepackage{xspace}
\usepackage[ruled, lined, linesnumbered, commentsnumbered, longend]{algorithm2e}
\usepackage{multirow,multicol}
\usepackage{colortbl}
\usepackage{caption}
\usepackage{tcolorbox}
\usepackage{mathtools}
\usepackage{tikz}
\usepackage{units}
\usepackage{listings}
\usepackage{ulem}
\usepackage{bbding}
\usepackage{bm}
\definecolor{backcolour}{rgb}{0.95,0.95,0.92}
\usepackage{enumitem}
\usepackage{graphicx}
\usepackage{subcaption}

\SetCommentSty{mycommfont}

\newcommand{\rqboxc}[1]{\begin{tcolorbox}[left=1pt,right=1pt,top=0pt,bottom=0pt,colback=gray!5,colframe=gray!40!black,before skip=5pt,after skip=0pt]#1\end{tcolorbox}}

\usepackage{tabularx}
\usepackage{listings}
\usepackage{color}

\definecolor{dkgreen}{rgb}{0,0.6,0}
\definecolor{gray}{rgb}{0.5,0.5,0.5}
\definecolor{mauve}{rgb}{0.58,0,0.82}

\usepackage{xcolor}

\graphicspath{ {./figures/} }
\usepackage{textgreek}
\usepackage{float}

\AtBeginDocument{%
  }
\begin{document}

\title{When LLMs Lag Behind: Knowledge Conflicts from Evolving APIs in Code Generation}

\author{Ahmed Nusayer Ashik}
\affiliation{%
  \institution{University of Manitoba}
  \city{Winnipeg}
  \country{Canada}
}
\email{ashikan@myumanitoba.ca}

\author{Shaowei Wang}
\affiliation{%
  \institution{University of Manitoba}
  \city{Winnipeg}
  \country{Canada}
}
\email{shaowei.wang@umanitoba.ca}

\author{Tse-Hsun Chen}
\affiliation{%
  \institution{Concordia University}
  \city{Montreal}
  \country{Canada}
}
\email{peterc@encs.concordia.ca}

\author{Muhammad Asaduzzaman}
\affiliation{%
  \institution{University of Windsor}
  \city{Windsor}
  \country{Canada}
}
\email{masaduzz@uwindsor.ca}

\author{Yuan Tian}
\affiliation{%
  \institution{Queen's University}
  \city{Kingston}
  \country{Canada}
}
\email{y.tian@queensu.ca}

\acmConference{}{}{}
\gdef\acmConference@shortname{}
\gdef\acmConference@date{}
\gdef\acmConference@venue{}
\makeatletter
\@ACM@journal@bibstriptrue
\makeatother




\begin{abstract}
The rapid evolution of software libraries creates a significant challenge for Large Language Models (LLMs), whose static parametric knowledge often becomes stale post-training. While retrieval-augmented generation (RAG) is commonly used to provide up-to-date API specifications, ``context-memory conflict'' arises when external instructions contradict a model's internal parametric knowledge. This paper presents a systematic empirical study of LLM code generation under API evolution (e.g., API deprecation, API modification, and API addition), by constructing a benchmark of 270 real-world updates from eight Python libraries. We evaluate four LLM families of 11 models. Our results show that without comprehensive documentation, LLMs struggle to prioritize external context, averaging only 42.55\% of generated code examples are executable in the target environment. While structured documentation and larger model scales improve LLMs' ability to update adoption, they do not fully resolve executability issues with a low 66.36\% executable rate. In addition, reasoning-based strategies (e.g., Self-Reflection) significantly boost LLMs' performance with 11\% improvement on executable rate. Our findings highlight the persistence of outdated patterns from LLMs, even when API update specifications are provided, and emphasize the need for evolution-aware benchmarks and techniques.

\end{abstract}



\maketitle

\fancyhead[LE]{}
\fancyhead[RO]{}


\section{Introduction}\label{sec:intro}

Software libraries are the backbone of modern software development. Developers routinely rely on well-established libraries to build complex, production-grade systems. Yet these libraries are far from static. Libraries evolve continuously (e.g., outdated Application Programming Interfaces (API) are deprecated or removed) to reflect improved design principles or performance requirements~\cite{xavier2017historical,lamothe2021systematic}. This continuous evolution creates a significant maintenance burden: developers must track changes, learn new idioms, and update their existing code base accordingly to avoid breaks.

Large language models (LLMs) have recently emerged as powerful coding assistants, demonstrating impressive ability to generate syntactically correct, contextually relevant code across a wide range of tasks, and are used to accelerate software development ~\cite{peng2023impact,ziegler2024measurin}. 
However, LLMs are fundamentally trained on static corpora with a fixed knowledge cutoff. The parametric knowledge encoded in their weights reflects the state of the world, and the state of library APIs as of their training data. When an API evolves after this cutoff, the model's internal knowledge becomes stale, and it may confidently generate code using outdated methods that no longer compile or produce correct behavior~\cite{versicode,llm_meet_lib_eval}.

A natural and widely adopted mitigation is retrieval-augmented generation (RAG) and prompt-based context injection~\cite{lewis2020retrieval, he2025evaluating,liang2025rustevo}. In software development contexts, this translates to developers supplying the LLM with up-to-date API specification (e.g., update description and updated API documentation) directly in the prompt, expecting the model to follow this external context and override its stale parametric knowledge. Yet this solution is deceptively fragile. LLMs are known to exhibit a fundamental tension between parametric knowledge (what they learned during training and store in parameters) and contextual knowledge (what is supplied at inference time)~\cite{xie2023adaptive,xu2024knowledge}. When provided context disagrees with a model's internal beliefs, context-memory conflict emerges. Recent research shows that models are often unable to fully suppress their internal knowledge even when explicitly instructed to do so, with performance consistently higher when contextual and parametric knowledge are aligned~\cite{sun2025seen,cheng2024understanding}. API evolution creates exactly such a conflict (i.e., old APIs in parametric knowledge vs. externally provided updated specification), it is deeply unclear how reliably models will defer to the provided context rather than fall back on memorized, outdated patterns.

To address this gap, we present the first systematic empirical study of LLM code generation under API evolution. We constructed a benchmark spanning real-world API updates across eight popular Python libraries and contains 270 updated API across three updated patterns (i.e., 45 deprecated/removed APIs, 128 modified APIs, and 97 newly introduced APIs). Those updated APIs are collected after the model knowledge cut-off date to ensure they do not exist in the model's parametric knowledge. We conducted our study on four LLM families (i.e., DeekSeek-Coder, CodeLlama, DeepSeek-R1-Qwen, and GPT-4o-mini) with varied sizes. Using the benchmark, we prompt LLMs to generate code examples that demonstrate correct usage of each updated API. To evaluate LLM performance, we design two complementary metrics. First, to measure whether the model acknowledges and responds to the externally provided update at the coarse level, we propose \textbf{API Adoption Rate} that measures whether the generated code at least partially adopt API update specification provided in the external context (e.g., invoking the recommended alternative in deprecated/removed APIs, even if not correctly), rather than ignoring them entirely. \textbf{Executable Rate}, computed exclusively over code examples that pass the first adoption filter, measures whether the generated code successfully runs in the target library environment, serving as the indicator of full and correct API usage. 

\begin{itemize}[leftmargin=*, itemsep=2pt, topsep=2pt]

    \item \textbf{RQ1:}  To what extent can LLMs follow externally provided API update specifications to generate correct and executable code? \textbf{Results:} Without full API documentation, LLMs fail to follow externally provided update specifications reliably — only 74.64\% adoption rate and 42.55\% executable rate on average. Adding structured API documentation yields significant improvement, yet executable rate remains low (66.36\%), and certain models (e.g., DeekSeek-R1) are even below 50\%.
    \item \textbf{RQ2:} How does performance vary across different model families and parameter scales? 
    \textbf{Results:} Different model families present varied behavior patterns. Scaling parameter size improves adoption but does not consistently resolve executability weaknesses. 
    \item \textbf{RQ3:} Can reasoning-based prompting techniques help LLMs follow the external context correctly?
    \textbf{Results:} Reasoning-based prompting (i.e., Chain-of-Thought (CoT) and Self-Reflection (SR)) yields marginal gain in adoption rate but substantial executability improvements (+11.33\% executable rate), with SR contributing more than CoT, particularly for API modification and addition patterns.
    \item \textbf{RQ4:} What errors arise when LLMs fail to follow external API update specifications?
    \textbf{Results:} LLM failures fall into two levels. At the adoption level, omission (42.1\%), old API usage (16.4\%), and hallucinated APIs dominate. At the execution level, over half of failures stem from incorrect API usage, such as incorrect parameters (26.6\%) and hallucinated behavior (16\%).
    
\end{itemize}

Our findings reveal certain implications, e.g., although providing API documentation can significantly improve the LLM's performance, certain models still struggle with a low executable rate. Reasoning-based prompt techniques, such as Self-Reflection, should be used to improve performance. The community needs continuously evolving benchmarks that track API changes beyond model training cutoffs to evaluate LLMs, explicitly accounting for post-training knowledge gaps and techniques to resolve them.

We make the following contributions:
\begin{itemize}[leftmargin=*, itemsep=2pt, topsep=2pt]
    \item We curate a benchmark of real-world API evolution across eight Python libraries, covering 270 updated APIs across three common patterns to evaluate the performance of LLMs on post-training knowledge that is not in their parametric knowledge.
    \item We conduct a systematic study on four LLM families (11 models) to measure their ability to generate code that follows externally provided API updates for code generation.
    \item We investigate two reasoning-oriented prompting strategies and show that they can improve LLM performance.
\end{itemize}

\section{Related Work}\label{sec:related}

\subsection{LLM-based code generation}

Many LLMs are developed specifically for coding-related tasks and are often referred to as code language models~\cite{jiang2026survey}, such as StarCoder~\cite{starcoder}, CodeLlama~\cite{code-llama}, Qwen2.5-Coder~\cite{qwen2.5-coder}, Deepseek-Coder~\cite{deepseekcoder}, and CodeGeeX~\cite{codegeex}. These models are trained on massive code corpora collected from software repositories, enabling strong performance across a wide range of programming tasks. 
With the rapid development of coding-oriented LLMs, a variety of approaches have been proposed, ranging from function-level synthesis~\cite{gu2025effectiveness,khojah2025impact,wu2024comprehensive,zan2024diffcoder} to repository-level code generation~\cite{zhang2024codeagent,li2025graphcodeagent,gao2025trae,yang2025lingxi,wang2024openhands}. For example, Zan et al. proposed DiffCoder, which improves function-level code generation involving API usage by modeling the differences (diffs) between related coding tasks, mimicking human analogical learning~\cite{zan2024diffcoder}. At the repository level, Zhang et al. introduced CODEAGENT, an LLM-based agent framework that integrates external tools to support complex, multi-file code generation tasks~\cite{codeapex}.
To systematically evaluate these models and approaches, a range of benchmark datasets have been developed~\cite{chen2024survey}. Representative function-level benchmarks include HumanEval~\cite{humaneval}, HumanEval+~\cite{humaneval+}, MBPP~\cite{mbpp}, EvalPlus~\cite{evalplus}, and APPS~\cite{apps}. For repository-level evaluation, commonly used benchmarks include SWE-bench~\cite{jimenez2023swe}, RepoBench~\cite{liu2023repobench}, CrossCodeEval~\cite{ding2023crosscodeeval}, and RepoFuse~\cite{liang2024repofuse}.

Despite these advances, existing studies either draw knowledge from models' parametric knowledge or predominantly assume that the knowledge required for code generation is consistent with the model’s parametric knowledge acquired during pre-training. In contrast, our work investigates an underexplored setting where external knowledge conflicts with the model’s internal knowledge. By systematically analyzing how LLMs prioritize and reconcile these conflicting sources, we reveal critical limitations in leveraging post-training knowledge and provide insights for improving model reliability in dynamically evolving software ecosystems.

\subsection{LLMs for Evolving APIs}

LLMs are typically trained on static, historical corpora, which inherently limits their ability to adapt to the evolutionary nature of software ecosystems. This temporal misalignment often leads to the generation of outdated or incorrect code. Prior studies have confirmed that deprecated API calls remain prevalent in LLM outputs; for instance, \citet{llm_meet_lib_eval} observed a 25\%-38\% deprecated API usage rate across eight Python libraries, attributing this failure to both stale parametric knowledge and the absence of real-time API status during inference.

Beyond simple deprecation, LLMs struggle with the nuances of version-specific evolution. Benchmarks such as GitChameleon 2.0 \cite{misra2025gitchameleon} reveal that even state-of-the-art models achieve only 48\%-51\% success in version-conditioned generation. Similarly, VersiCode \cite{versicode} and LibEvolutionEval \cite{kuhar2025libevolutioneval} demonstrate that models frequently fail to adhere to version-specific instructions, showing significant performance volatility as APIs evolve. While Retrieval-Augmented Generation (RAG) is the standard mitigation—improving success rates by up to 13.5\% on unseen Rust APIs \cite{liang2025rustevo}, providing the correct documentation does not guarantee the model will follow it. 
Although providing the updated API specification is a natural way to mitigate the issue, it is unclear if LLMs really follow the externally provided information, and to what extent it follows, typically when the internal knowledge and externally provided knowledge are in conflict. Our study fills this gap. In addition, we provide the first systematic evaluation of how reasoning-based techniques, such as Self-Reflection and Chain-of-Thought, serve as mechanisms to resolve these internal knowledge conflicts.



\section{Methodology}\label{sec:method}

\begin{figure*}
    \centering
    \includegraphics[width=1\linewidth]{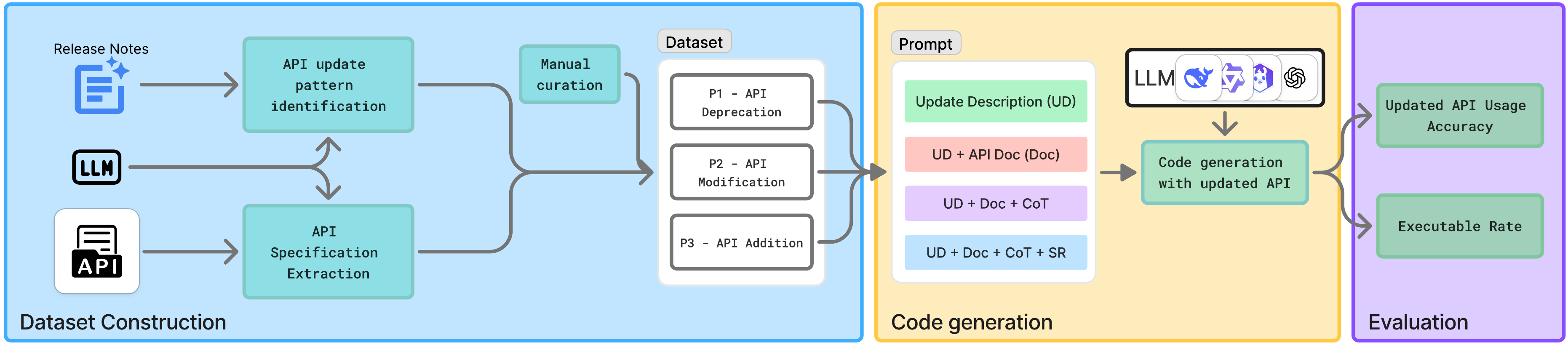}
      \vspace{-0.2in}
    \caption{The overall workflow of our study.}
    \label{fig:workflow}
      \vspace{-0.2in}
\end{figure*}

Figure~\ref{fig:workflow} presents the overall workflow of our study, which consists of three main phases: (1) constructing a dataset of updated APIs and their associated metadata, (2) prompting the studied LLMs to generate code examples demonstrating the usage of these updated APIs, and (3) evaluating the quality of the generated code examples. We describe each phase in detail below.

\subsection{Dataset Construction}
To rigorously assess whether LLMs can correctly leverage updated API knowledge provided in the external context, we construct a benchmark dataset of API changes introduced after the knowledge cutoff of the target LLMs. Our construction process proceeds in the following steps.

\subsubsection{Library and Version Selection}
To ensure LLMs know the APIs, we target eight popular and widely used Python libraries: NumPy, Pandas, scikit-learn, SciPy, JAX, Keras, TensorFlow, and PyTorch. These libraries are widely adopted in data processing, machine learning, and deep learning, and have been commonly used in prior studies on API evolution~\cite{wang2025llms, wang2020exploring}. Since our goal is to evaluate LLMs' ability to follow externally provided API knowledge rather than recall memorized patterns, we restrict our dataset to API changes introduced in versions released after December 2023, beyond the knowledge cutoff of most widely used LLMs~\cite{deepseekcoder,gpt4o-mini,code-llama,guo2025deepseek}. This ensures that the API changes in our dataset are unlikely to be encoded in the models' parametric knowledge, isolating the effect of external context provision. For each library, we identify all versions released after this cutoff date and collect their associated release notes from the corresponding GitHub repositories.

\subsubsection{API Update Pattern Identification}
Release notes serve as the primary source for identifying API changes, as they systematically summarize all modifications introduced in a given version, including bug fixes, behavioral changes, and API updates. In this study, we focus on three common API update patterns~\cite{zhang2020python}: 
\begin{itemize}
    \item \textbf{P1 - API Deprecation}: APIs that are removed or deprecated, typically with a recommended alternative one. Note that in this study, we only focus on the deprecated APIs that have alternative ones suggested, since our goal is to generate code examples to adopt the updates.
    \item \textbf{P2 - API Modification}: APIs whose function name, parameters, or return type have been changed.
    \item \textbf{P3 - API Addition}: Newly introduced APIs that did not exist in prior versions.
\end{itemize}

Manually extracting and classifying API changes from release notes at scale is labor-intensive and error-prone. To address this, we leverage an LLM to automate the extraction process, taking advantage of its strong capabilities in natural language understanding and structured information extraction from software documentation~\cite{bhattacharyya2025information, daneshyan2025smartnote}. Specifically, we prompt the LLM to parse each release note, identify API changes, and classify them into one of the three update patterns. For each extracted API change, we record: (1) the library name, (2) the version introducing the change, (3) the updated function name, and (4) a textual description of the update. We use GPT-5 mini~\cite{singh2025openai} for this extraction step, given its strong instruction-following ability and cost efficiency at scale.

\subsubsection{API Specification Extraction}
For each identified API change, we further collect the corresponding API documentation from the library's official documentation pages. Specifically, we extract function signatures, parameter descriptions, return type information, code usage examples, and any additional notes available in the documentation. This documentation serves as the external context that will be provided to LLMs during code generation.
Entries for which no official documentation could be located are removed from the dataset to ensure the completeness and quality of the context provided to models. For P1 (API Deprecation) instances, our evaluation focuses on whether LLMs can generate correct code using the \emph{alternative} API recommended to replace the deprecated or removed one, since the goal is to assess whether models can migrate away from stale API usage given external guidance. Accordingly, we also collect documentation for each recommended alternative API and discard P1 instances where alternative API documentation is unavailable. Similarly, we use LLM to extract API specification information from API documentation.  

\subsubsection{Manual curation}
To ensure dataset quality, two authors independently inspected all collected API change instances and removed entries that did not meet our quality criteria. Specifically, we discarded instances where: (1) the extracted API change was incorrectly classified or did not correspond to a genuine API update (e.g., false positives introduced by the LLM-based extraction); (2) the API documentation was incomplete, ambiguous, or missing critical information such as function signatures or parameter descriptions. Any disagreements between the two authors during inspection were resolved through discussion until consensus was reached. This manual curation step ensures that every instance in our final dataset provides a well-defined, complete, and unambiguous specification that can serve as a reliable external context for LLM code generation.

\subsubsection{Resulting Dataset}
The dataset has 270 data instances across 3 patterns. The details of each pattern are presented in Table~\ref{tab:library-patterns}.

\begin{table}[h]
\caption{Data statistics for the three API update patterns, i.e., P1 - API Deprecation, P2- API Modification, and P3-API Addition}
\label{tab:library-patterns}
\resizebox{\columnwidth}{!}{%
\begin{tabular}{lp{1in}p{0.5in}p{0.5in}p{0.5in}}
\toprule
\textbf{Library Name} & \textbf{Versions} & \textbf{P1} & \textbf{P2} & \textbf{P3} \\
\midrule

Numpy       & 2.0.0 -- 2.2.0    & 17  & 9   & 18 \\ \hline
Pandas      & 2.2.0 -- 2.2.1    & 3  & 10  & 2  \\ \hline
Scipy       & 1.12.0 -- 1.15.0  & 19  & 49  & 50 \\ \hline
Scikit Learn & 1.4 -- 1.6       & 1   & 30  & 4  \\ \hline
Matplotlib  & 3.9.0 -- 3.10.0   & --  & 13  & 7  \\ \hline
Keras       & 3.10.0            & --  & 12  & -- \\ \hline

\multirow{5}{*}{Jax} 
            & 0.4.28 -- 0.4.29      & \multirow{5}{*}{4} & \multirow{5}{*}{5} & \multirow{5}{*}{16} \\
            & 0.4.31 -- 0.4.32  &    &    &    \\
            & 0.4.34 -- 0.4.36  &    &    &    \\
            & 0.4.38  &    &    &    \\
            & 0.5.0 -- 0.5.1           &    &    &    \\ 
            & 0.6.0 -- 0.6.2           &    &    &    \\
            \hline

Pytorch     & 2.7.0             & 1   & --  & -- \\ 
\midrule
\textbf{Total}  &  & \textbf{45} & \textbf{128} & \textbf{97} \\
\bottomrule
\end{tabular}%
}
\vspace{-0.2in}
\end{table}

\subsection{Code Generation}\label{sec:task}

To simulate a realistic developer workflow, we design a code generation task for each API instance in our benchmark. Specifically, we prompt each LLM to generate a self-contained code example that correctly demonstrates the usage of the target updated API. Each prompt supplies two forms of external context: (1) an \textit{update description (UD)} summarizing the nature of the API change (e.g., \textit{The function scipy.histogram is deprecated and removed from SciPy's main namespace. Use numpy.histogram directly.}), and (2) the full \textit{API documentation} (Doc) for the updated API, including its signature, parameters, and usage notes. Together, these mirror the information a developer would consult when migrating to or adopting an updated API.
The task is intentionally designed to create a conflict between external context and parametric knowledge: the correct implementation requires the model to follow the updated API specification provided in the prompt, which may directly contradict the outdated API patterns encoded in the model's weights. 

\begin{figure}[t]
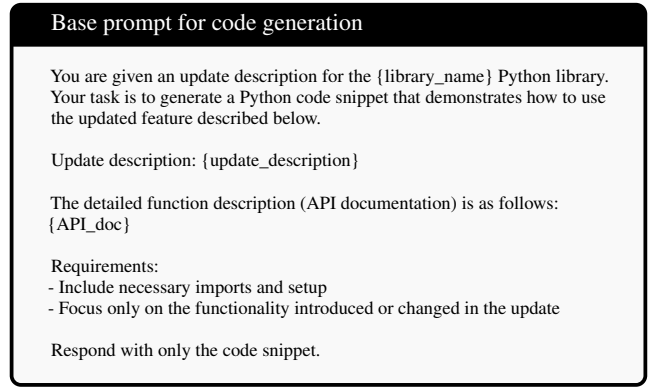

\centering
\begin{tcolorbox}[
    title=Base prompt for code generation,
    colback=gray!5,
    colframe=black
]
\footnotesize
You are given an update description for the \{library\_name\} Python library. Your task is to generate a Python code snippet that demonstrates how to use the updated feature described below. \\

Update description: \{update\_description\} \\

The detailed function description (API documentation) is as follows: \\
\{API\_doc\} \\

Requirements: \\
- Include necessary imports and setup \\
- Focus only on the functionality introduced or changed in the update \\

Respond with only the code snippet.
\end{tcolorbox}
\vspace{-0.2in}
\caption{Base prompt for code generation}
\label{fig:base_prompt}
\vspace{-0.3in}
\end{figure}

Prior work has shown that clearly structured external context can strongly influence LLM responses even when it conflicts with parametric knowledge~\cite{xie2023adaptive}; our task design allows us to directly measure the extent to which this holds for API-level code generation.
The prompt structure varies slightly by update pattern. For \textbf{P2 (API Modification)}, the model is asked to generate code using the modified API as specified in the provided documentation. For \textbf{P1 (API Deprecation)}, the model is prompted to generate code using the recommended alternative API described in the update. For \textbf{P3 (API Addition)}, the model is asked to generate code demonstrating the usage of the newly introduced API. Figure~\ref{fig:base_prompt} illustrates the prompt template for P2; templates for P1 and P3 follow the same structure with pattern-appropriate instructions.
\\
\\
To isolate the contribution of the API documentation to model performance, we additionally construct a \textit{reduced-context} condition in which the model receives only the update description (UD) without the full API documentation. Comparing performance between the full-context condition (UD + Doc) and the reduced-context condition (UD only) allows us to quantify how much the structured API documentation contributes to correct code generation.

\subsubsection{Reasoning-based prompting techniques}
Beyond standard prompting, we investigate whether reasoning-based prompting techniques can help LLMs better follow externally provided API context and reduce their reliance on outdated parametric knowledge. Prior work has shown that reasoning-oriented prompting encourages more deliberate, structured analysis before producing a final output~\cite{liu2023pre, sahoo2024systematic}, which may help models more faithfully interpret and apply updated API specifications. We evaluate two widely adopted techniques that represent complementary reasoning strategies: one that augments the generation process with explicit intermediate reasoning, and one that introduces a post-generation verification and correction step.

\noindent\textbf{Chain-of-Thought (CoT)}~\cite{wei2022chain} prompts the model to explicitly reason through the problem before producing the final code. Rather than generating code directly, the model is first asked to analyze the provided update description and API documentation before the implementation. This explicit reasoning step is intended to direct the model's attention toward the external context and surface potential conflicts with its parametric knowledge before they manifest as errors in the generated code. 




\noindent\textbf{Self-Reflection (SR)}~\cite{madaan2023self} introduces a two-stage generation process in which the model first produces an initial code example and then critically reviews its own output. In the reflection stage, the model is explicitly asked to examine whether the generated implementation correctly follows the provided API documentation and update description, identify any inconsistencies — particularly cases where outdated API patterns may have been used — and revise the code accordingly. Unlike CoT, which intervenes \textit{before} generation, SR operates \textit{after} an initial attempt, making it complementary as a post-hoc correction mechanism. We append the self-reflection instruction as presented below. 

\begin{tcolorbox}[title=Prompt template for Self-Reflection, colback=gray!5, colframe=black]
\footnotesize
... \\

Self-Verification Step:\\
- After generating the code snippet, verify that:\\
\hspace*{1em}1) The deprecated/removed function does NOT appear in the code.\\
\hspace*{1em}2) The alternate function is correctly used with all required parameters.\\
\hspace*{1em}3) No unsupported or invented arguments are used.\\
\hspace*{1em}4) The code is syntactically valid Python and runnable.\\
- If any check fails, regenerate the code snippet and re-verify.\\
- Output only the final verified Python code snippet.
\end{tcolorbox}

\subsection{Evaluation}

We evaluate the quality of the generated code along two complementary metrics. 

\noindent\textbf{API Adoption Rate}, measures whether the generated code shows any attempt to adopt the updated API specification provided in the external context: that is, whether the model \textit{at least partially} incorporates the API changes described in the prompt rather than completely ignoring them. For P1 (API Deprecation), this means the generated code makes at least some use of the recommended alternative API, even if it is not used correctly (i.e., wrong parameters), rather than continuing to invoke the removed/deprecated one or use other irrelevant APIs. For P2 (API Modification), where the changes are typically minimal, often involving only a single updated parameter, this means the generated code reflects at least some aspect of the modification, such as using the revised parameter structure, even if the parameter positioning or data type is not entirely correct. For P3 (API Addition), similar to P1, this means the generated code attempts to use the newly introduced API, even though it is not complete. This metric aims to measure how prevalent the model completely omits the provided external context. API Adoption Rate serves as a coarse but critical first filter, establishing the baseline of whether the model at least acknowledges and responds to the externally provided API update.
\[
API Adoption Rate = \frac{\text{\#Partially or fully adopted code examples}}{\text{\# Total code examples}}
\]

To scale evaluation across our full benchmark, we adopt an LLM-as-a-Judge approach~\cite{zheng2023judging}, in which a judge model automatically determines whether each generated code example correctly follows the updated API specification provided in the prompt. The judge receives the update description, the API documentation, and the generated code as input, and is queried to produce a binary output (i.e., at least partially adopt or not). We use GPT5-mini as the judge model. To validate the reliability of this automated evaluation, two authors manually annotated a randomly sampled set of 1,000 generated code examples and compared these annotations against the LLM judge's decisions. The judge achieved a labeling accuracy of 91.4\% against the manual annotations, confirming that the automated evaluation is sufficiently reliable for large-scale analysis.

\noindent\textbf{Executable Rate.}
Since static inspection in API Adoption Rate alone is insufficient to determine whether an API update has been completely and correctly adopted, we execute each generated code example in the target environment and use successful execution as the indicator of full correctness. This metric is computed exclusively over the subset of generated code examples that have already been identified as at least partially adopting the updated API, i.e., those that passed the API Adoption Rate filter. This restriction is intentional, since including code examples that show no adoption of the update would introduce bias, as such examples may still execute successfully by using entirely different, unrelated APIs that happen to be valid in the target environment.

Code examples are extracted from model responses using regular expressions. Each example is then executed in an isolated virtual environment configured with the exact library version in which the corresponding API change was introduced, as recorded in our benchmark. Using the precise target version ensures that the execution environment faithfully reflects the state of the library at the point of the API change. Each virtual environment contains only the required library and its dependencies, preventing interference from other installed packages and eliminating version contamination that could otherwise introduce false positives or false negatives in the execution results.The Executable Rate is computed as:

\[
ExecutableRate = \frac{\text{\#Executable code examples}}{\text{\#Partially or fully adopted code examples}}
\]

\section{Experimental Setting}\label{sec:experimentalsetting}

\subsection{Research Questions}

\begin{itemize}[leftmargin=*, itemsep=2pt, topsep=2pt]
    \item \textbf{RQ1:} To what extent can LLMs follow externally provided API update specifications to generate correct and executable code?
    
    \item \textbf{RQ2:} How does performance vary across different model families and parameter scales?
    
    \item \textbf{RQ3:} Can reasoning-based prompting techniques help LLMs to follow the external context correctly?
    

    \item \textbf{RQ4:} What errors arise when LLMs fail to follow external API update specifications?

\end{itemize}

\subsection{Evaluated LLMs}

To evaluate how different LLMs handle evolved APIs that are not stored in their parametric knowledge, we select models from multiple model families and parameter scales. Our selection is guided by four criteria: 1) The models must have a knowledge cutoff prior to December 2023, ensuring that the API updates collected in our benchmark (introduced after December 2023) are unlikely to be included in their training data. This requirement enables us to reliably construct scenarios where the models possess outdated parametric knowledge of the APIs. 2) We select models that provide multiple parameter scales within the same model family. This allows us to investigate how knowledge conflict behavior varies with model size. 3) We include both open-source and proprietary models to ensure the generalizability of our findings across different model ecosystems. 4) We include both reasoning and non-reasoning models.

Following these criteria, we select four representative model series and Table~\ref{tab:models} summarizes the selected models, their parameter sizes, knowledge cutoff dates, and open-source availability.

\begin{table}[h]
  \caption{Summary of evaluated LLMs.}
    \vspace{-0.2in}
  \label{tab:models}
  \resizebox{\columnwidth}{!}{%
  \begin{tabular}{lllc}
    \toprule
    Model Series & Size Variants & Cutoff Date & Open Source \\
    \midrule
    DeepSeek-Coder~\cite{guo2024deepseek} & 1.3B, 6.7B, 33B & Before Nov 2023  & \ding{51} \\
    CodeLlama~\cite{code-llama} & 7B, 13B, 34B & \makecell[l]{No official date\\(Released on Aug 2023)} & \ding{51} \\
    DeepSeek-R1-Qwen~\cite{guo2025deepseek} & 1.5B, 7B, 14B, 32B & \makecell[l]{Late 2023} & \ding{51} \\
    GPT-4o mini~\cite{gpt4o-mini} & N/A & Oct 2023 & \ding{55} \\
    \bottomrule
  \end{tabular}%
  }
  \vspace{-0.2in}
\end{table}

\subsection{Approaches of RQs}

To answer RQ1 and RQ2, we conduct controlled experiments using the code generation task described in Section~\ref{sec:task} across all API instances in our benchmark and all selected LLMs. For each API instance, we prompt the LLM to generate a code example using the update description and the corresponding API documentation. We collect the generated code examples and measure the Accuracy for each pattern across all studied LLMs. 

To answer RQ3, we compare the baseline (UD+Doc) where only the update description and API documentation are provided, with the prompts with CoT and SR (i.e., UD+Doc+CoT and UD+Doc+CoT+SR). All prompts contain the same external context, including the API update description and the corresponding documentation. The only difference between configurations is the presence of reasoning instructions. For each configuration, we generate code for all API instances using the evaluated LLMs and measure their accuracy and executable rate. 

To answer RQ4, we analyzed the failure cases from all models that still exist under the best configuration (UD+Doc+CoT+SR) in our study (see results in Section~\ref{sec:rq3}). Specifically, we analyzed two levels of failures. \textbf{L1:} all cases where the LLM failed to adopt the API update specification at all (195 cases in total); and \textbf{L2:} all cases where the update was at least partially adopted but the generated code failed to execute (781 cases in total). For L1, we first run a pilot study on randomly sampled failure cases to derive the following failure categories:
\begin{itemize}
    \item \textbf{Hallucination:} Generated code that is not relevant to the updates
    described in the update description.
    \item \textbf{Omission:} Completely ignored requirements specified in the update (e.g., API/Alternate API in P1 and P3, and updated parameters in P2).
    \item \textbf{Old API Usage:} Used only the deprecated or removed API that the update is meant to replace. The new/alternate API does not appear in the code.
    \item \textbf{Mixed API Usage:} Used both old and new API patterns in the same snippet.
    \item \textbf{Others:} Does not fit any of the above.
\end{itemize}

To scale up our failure analysis to all failure cases, we used GPT-5 mini as a code review judge to classify the failure cases in to above derived categories. We provided the judge LLM with the API update description, the LLM-generated code in response to that update, and the definition of the failure categories, and asked it to identify in which category the failure falls.

For L2,  following a similar process as 1), we first derive the following failure categories and then apply LLM-as-a-judge to classify the failure category for each case.
\begin{itemize}
    
    \item \textbf{Hallucinated Behavior}: Used the API update but it behaves in a way that contradicts its actual specification (e.g., assumed a return type, chained a method that does not exist on the result). 
    \item \textbf{Wrong Parameters:} Used the API update but with incorrect arguments, wrong parameter names, wrong types, or wrong argument order. 
    \item \textbf{Incompatible Context:} Used the updated API with correct syntax but in
a context where it cannot work (e.g., initialization of the data is wrong). 
    \item \textbf{Missing Setup}: Adopted the API update correctly in isolation but missed required setup steps such as imports and configuration of the environment. 
    \item \textbf{Not Update Related:} The error is unrelated to the API usage (e.g., a typo and an unrelated syntax error to the updated API).
\end{itemize}

To ensure the reliability of our annotation, we manually labeled 50 failure cases (20 cases from L1, and 30 cases from L2. Accuracy is 90\% (45/50).


\subsection{Implementation Details} 

All experiments are done in Python 3.10. We used 8 bit quantized model for Deepseek-coder-33B, CodeLlama-34B, and Deepseek-r1-qwen-33B. All the experiments were conducted on a server with four RTX 3090 GPUs and 256G RAM.

\section{Results}\label{sec:results}
\subsection{RQ1 - API update specification compliance}

\newcommand{\rot}[1]{\rotatebox[origin=l]{90}{\makebox[2.3cm][l]{\small #1}}}

\newcommand{\cellone}[2][white]{%
  \cellcolor{#1}#2\%%
}

\definecolor{green1}{HTML}{1B7837}  
\definecolor{green2}{HTML}{5AAE61}  
\definecolor{green3}{HTML}{A6D96A}  
\definecolor{green4}{HTML}{D9F0A3}  
\definecolor{green5}{HTML}{F7FCB1}  

\definecolor{lowcolor}{HTML}{f9fbe6}
\definecolor{highcolor}{HTML}{38ab5b}

\newcommand{\minvalupdate}{40.63}
\newcommand{\maxvalupdate}{98.43}
\newcommand{\midvalupdate}{89.00}

\newcommand{\minvaladd}{60.82}
\newcommand{\maxvaladd}{100}
\newcommand{\midvaladd}{98.97}

\newcommand{\minvaldelete}{15.56}
\newcommand{\maxvaldelete}{100}
\newcommand{\midvaldelete}{97.00}

\definecolor{color1}{HTML}{FFFFF0}   
\definecolor{color2}{HTML}{A8D8A8}   
\definecolor{color3}{HTML}{5BBD6F}   
\definecolor{color4}{HTML}{1B8B4B}   

\newcommand{\celladd}[1]{%
  \ifdim#1pt<\midvaladd pt
    \pgfmathparse{(\minvaladd == \midvaladd) ? 50 : (#1 - \minvaladd) / (\midvaladd - \minvaladd) * 100}%
    \xdef\mix{\pgfmathresult}%
    \cellcolor{color2!\mix!color1}#1\%
  \else
    \pgfmathparse{(\midvaladd == \maxvaladd) ? 50 : (#1 - \midvaladd) / (\maxvaladd - \midvaladd) * 100}%
    \xdef\mix{\pgfmathresult}%
    \cellcolor{color4!\mix!color3}#1\%
  \fi
}

\newcommand{\cellupdate}[1]{%
  \ifdim#1pt<\midvalupdate pt
    \pgfmathparse{(\minvalupdate == \midvalupdate) ? 50 : (#1 - \minvalupdate) / (\midvalupdate - \minvalupdate) * 100}%
    \xdef\mix{\pgfmathresult}%
    \cellcolor{color2!\mix!color1}#1\%
  \else
    \pgfmathparse{(\midvalupdate == \maxvalupdate) ? 50 : (#1 - \midvalupdate) / (\maxvalupdate - \midvalupdate) * 100}%
    \xdef\mix{\pgfmathresult}%
    \cellcolor{color4!\mix!color3}#1\%
  \fi
}

\newcommand{\celldelete}[1]{%
  \ifdim#1pt<\midvaldelete pt
    \pgfmathparse{(\minvaldelete == \midvaldelete) ? 50 : (#1 - \minvaldelete) / (\midvaldelete - \minvaldelete) * 100}%
    \xdef\mix{\pgfmathresult}%
    \cellcolor{color2!\mix!color1}#1\%
  \else
    \pgfmathparse{(\midvaldelete == \maxvaldelete) ? 50 : (#1 - \midvaldelete) / (\maxvaldelete - \midvaldelete) * 100}%
    \xdef\mix{\pgfmathresult}%
    \cellcolor{color4!\mix!color3}#1\%
  \fi
}

\begin{table*}[t]
\centering
\caption{The API adoption rate across different API update patterns and prompting strategies. A larger value is preferable and is marked in a darker color.}
\label{tab:results_adoption_rate}
\setlength{\tabcolsep}{4.8pt}
\renewcommand{\arraystretch}{1.1}
\footnotesize
\begin{tabular}{@{}l cccc | cccc | cccc @{}}
\toprule
 & \multicolumn{4}{c}{\textbf{P1 -- API Deprecation}}
 & \multicolumn{4}{c}{\textbf{P2 -- API Modification}}
 & \multicolumn{4}{c}{\textbf{P3 -- API Addition}} \\
\cmidrule(lr){2-5} \cmidrule(lr){6-9} \cmidrule(l){10-13}
\textbf{LLM}
 & UD
 & UD+Doc
 & +CoT
 & +CoT/SR
  & UD
 & UD+Doc
 & +CoT
 & +CoT/SR
  & UD
 & UD+Doc
 & +CoT
 & +CoT/SR\\
\midrule
Deepseek-coder-1.3B-instruct  
  & \celldelete{75.56}
  & \celldelete{97.78}
  & \celldelete{97.78}
  & \celldelete{100}
  & \cellupdate{47.66}
  & \cellupdate{71.87}
  & \cellupdate{64.06}
  & \cellupdate{69.53}
  & \celladd{94.85}
  & \celladd{98.97}
  & \celladd{100}
  & \celladd{98.97} \\
  
Deepseek-coder-6.7B-instruct  
  & \celldelete{66.67}
  & \celldelete{100}
  & \celldelete{97.78}
  & \celldelete{100}
  & \cellupdate{64.06}
  & \cellupdate{82.03}
  & \cellupdate{85.94}
  & \cellupdate{86.72}
  & \celladd{73.2}
  & \celladd{90.72}
  & \celladd{96.91}
  & \celladd{95.88} \\

Deepseek-coder-33B-instruct  
  & \celldelete{73.33}
  & \celldelete{97.78}
  & \celldelete{100}
  & \celldelete{100}
  & \cellupdate{82.03}
  & \cellupdate{90.62}
  & \cellupdate{92.97}
  & \cellupdate{94.53}
  & \celladd{68.04}
  & \celladd{95.88}
  & \celladd{98.97}
  & \celladd{98.97} \\
  \hline

CodeLlama-7B-Instruct  
  & \celldelete{20.00}
  & \celldelete{86.67}
  & \celldelete{100}
  & \celldelete{97.78}
  & \cellupdate{72.66}
  & \cellupdate{89.06}
  & \cellupdate{92.97}
  & \cellupdate{89.06}
  & \celladd{97.94}
  & \celladd{100}
  & \celladd{100}
  & \celladd{100} \\

CodeLlama-13B-Instruct  
  & \celldelete{15.56}
  & \celldelete{88.89}
  & \celldelete{95.56}
  & \celldelete{100}
  & \cellupdate{82.03}
  & \cellupdate{92.97}
  & \cellupdate{92.19}
  & \cellupdate{93.75}
  & \celladd{100}
  & \celladd{100}
  & \celladd{98.97}
  & \celladd{100} \\

CodeLlama-34B-Instruct
  & \celldelete{88.89}
  & \celldelete{97.78}
  & \celldelete{97.78}
  & \celldelete{100}
  & \cellupdate{84.37}
  & \cellupdate{92.19}
  & \cellupdate{95.31}
  & \cellupdate{89.06}
  & \celladd{97.94}
  & \celladd{98.97}
  & \celladd{100}
  & \celladd{100} \\
\hline

DeepSeek-R1-Distill-Qwen-1.5B 
  & \celldelete{44.44}
  & \celldelete{91.11}
  & \celldelete{86.67}
  & \celldelete{95.56}
  & \cellupdate{60.94}
  & \cellupdate{71.09}
  & \cellupdate{72.66}
  & \cellupdate{68.75}
  & \celladd{74.23}
  & \celladd{82.47}
  & \celladd{88.67}
  & \celladd{89.69} \\

DeepSeek-R1-Distill-Qwen-7B 
  & \celldelete{48.89}
  & \celldelete{95.56}
  & \celldelete{97.78}
  & \celldelete{100}
  & \cellupdate{83.59}
  & \cellupdate{90.62}
  & \cellupdate{91.41}
  & \cellupdate{89.06}
  & \celladd{92.78}
  & \celladd{98.97}
  & \celladd{94.84}
  & \celladd{98.97} \\

DeepSeek-R1-Distill-Qwen-14B 
  & \celldelete{60.00}
  & \celldelete{95.56}
  & \celldelete{97.78}
  & \celldelete{100}
  & \cellupdate{86.72}
  & \cellupdate{91.41}
  & \cellupdate{92.19}
  & \cellupdate{92.19}
  & \celladd{96.91}
  & \celladd{98.97}
  & \celladd{98.97}
  & \celladd{98.97} \\

DeepSeek-R1-Distill-Qwen-32B  
  & \celldelete{68.89}
  & \celldelete{91.11}
  & \celldelete{100}
  & \celldelete{100}
  & \cellupdate{87.50}
  & \cellupdate{89.84}
  & \cellupdate{93.75}
  & \cellupdate{94.53}
  & \celladd{94.84}
  & \celladd{100}
  & \celladd{95.88}
  & \celladd{98.97} \\
\hline
GPT-4o mini 
  & \celldelete{88.89}
  & \celldelete{100}
  & \celldelete{100}
  & \celldelete{100}
  & \cellupdate{94.53}
  & \cellupdate{96.87}
  & \cellupdate{96.09}
  & \cellupdate{97.66}
  & \celladd{77.32}
  & \celladd{98.97}
  & \celladd{98.97}
  & \celladd{100} \\
 \bottomrule
\end{tabular}

\smallskip
\raggedright
\footnotesize
\textbf{Legend:} UD = Update Description; UD+Doc = Update Description + API Documentation; +CoT = UD+Doc+CoT; +CoT/SR = UD+Doc+CoT+SR.
\end{table*}


\newcommand{\cell}[4][white]{%
  \begin{tabular}{@{}c@{}}
    \cellcolor{#1} #2\% \\[-2pt]
    {\scriptsize (#3/#4)}
  \end{tabular}%
}

\newcommand{\minvaladderror}{19.59}
\newcommand{\maxvaladderror}{78.82}

\definecolor{colorlow}{HTML}{098339}    
\definecolor{colorhigh}{HTML}{FFFFD4}   

\newcommand{\celladderror}[3]{%
  \pgfmathparse{(\minvaladderror == \maxvaladderror) ? 50 : (#1 - \minvaladderror) / (\maxvaladderror - \minvaladderror) * 100}%
  \pgfmathparse{min(max(\pgfmathresult, 0), 100)}%
  \xdef\mix{\pgfmathresult}%
  \begin{tabular}{@{}c@{}}
    \cellcolor{colorlow!\mix!colorhigh} #1\% \\[-2pt]
    {\scriptsize (#2/#3)}
  \end{tabular}%
}


\begin{table*}[t]
\centering
\caption{Executable Rate of generated code examples that follow the updated APIs across models and prompting strategies. A larger value indicates a higher executable rate and is marked with a darker color.}
\label{tab:ExecRate_results}
\setlength{\tabcolsep}{4.8pt}
\renewcommand{\arraystretch}{1.1}
\footnotesize
\begin{tabular}{@{}l cccc | cccc | cccc @{}}
\toprule
  & \multicolumn{4}{c}{\textbf{P1 -- API Deprecation}}
 & \multicolumn{4}{c}{\textbf{P2 -- API Modification}}
 & \multicolumn{4}{c}{\textbf{P3 -- API Addition}} \\
\cmidrule(lr){2-5} \cmidrule(lr){6-9} \cmidrule(l){10-13}
\textbf{LLM}
 & UD
 & UD+Doc
 & +CoT
 & +CoT/SR
  & UD
 & UD+Doc
 & +CoT
 & +CoT/SR
  & UD
 & UD+Doc
 & +CoT
 & +CoT/SR\\
\midrule
Deepseek-coder-1.3B-instruct  
  & \celladderror{50.00}{17}{34}
  & \celladderror{77.27}{34}{44}
  & \celladderror{68.18}{30}{44}
  & \celladderror{66.67}{30}{45}
  & \celladderror{34.43}{21}{61}
  & \celladderror{63.04}{58}{92}
  & \celladderror{67.07}{55}{82}
  & \celladderror{70.79}{63}{89}
  & \celladderror{38.04}{35}{92}
  & \celladderror{78.12}{75}{96}
  & \celladderror{80.41}{78}{97}
  & \celladderror{75.00}{72}{96} \\
  
Deepseek-coder-6.7B-instruct
  & \celladderror{52.94}{16}{34}
  & \celladderror{86.67}{39}{45}
  & \celladderror{88.64}{39}{44}
  & \celladderror{82.22}{37}{45}
  & \celladderror{41.46}{34}{82}
  & \celladderror{73.33}{77}{105}
  & \celladderror{70.91}{78}{110}
  & \celladderror{75.68}{84}{111}
  & \celladderror{43.48}{30}{69}
  & \celladderror{77.27}{68}{88}
  & \celladderror{80.85}{76}{94}
  & \celladderror{80.65}{75}{93} \\
  
Deepseek-coder-33B-instruct
  & \celladderror{66.67}{22}{33}
  & \celladderror{79.55}{35}{44}
  & \celladderror{84.44}{38}{45}
  & \celladderror{84.44}{38}{45}
  & \celladderror{62.86}{66}{105}
  & \celladderror{68.10}{79}{116}
  & \celladderror{64.71}{77}{119}
  & \celladderror{71.90}{87}{121}
  & \celladderror{39.39}{26}{66}
  & \celladderror{78.49}{73}{93}
  & \celladderror{77.89}{74}{95}
  & \celladderror{78.12}{75}{96} \\
  \hline
  
CodeLlama-7B-Instruct
  & \celladderror{11.11}{1}{9}
  & \celladderror{66.67}{26}{39}
  & \celladderror{86.67}{39}{45}
  & \celladderror{75.00}{33}{44}
  & \celladderror{27.96}{26}{93}
  & \celladderror{59.65}{68}{114}
  & \celladderror{61.34}{73}{119}
  & \celladderror{70.18}{80}{114}
  & \celladderror{41.05}{39}{95}
  & \celladderror{75.26}{73}{97}
  & \celladderror{78.35}{76}{97}
  & \celladderror{78.35}{76}{97} \\
  
CodeLlama-13B-Instruct
  & \celladderror{14.29}{1}{7}
  & \celladderror{65.00}{26}{40}
  & \celladderror{67.44}{29}{43}
  & \celladderror{68.89}{31}{45}
  & \celladderror{41.90}{44}{105}
  & \celladderror{14.29}{17}{119}
  & \celladderror{30.77}{36}{117}
  & \celladderror{64.17}{77}{120}
  & \celladderror{49.48}{48}{97}
  & \celladderror{71.13}{69}{97}
  & \celladderror{68.75}{66}{96}
  & \celladderror{72.16}{70}{97} \\
  
CodeLlama-34B-Instruct
  & \celladderror{40.00}{16}{40}
  & \celladderror{65.91}{29}{44}
  & \celladderror{56.82}{25}{44}
  & \celladderror{77.78}{35}{45}
  & \celladderror{47.22}{51}{108}
  & \celladderror{63.37}{64}{101}
  & \celladderror{64.15}{68}{106}
  & \celladderror{70.18}{80}{114}
  & \celladderror{45.26}{43}{95}
  & \celladderror{72.92}{70}{96}
  & \celladderror{76.29}{74}{97}
  & \celladderror{73.20}{71}{97} \\
  \hline

DeepSeek-R1-Distill-Qwen-1.5B
  & \celladderror{30.00}{6}{20}
  & \celladderror{66.67}{26}{39}
  & \celladderror{86.84}{33}{38}
  & \celladderror{76.74}{33}{43}
  & \celladderror{22.37}{17}{76}
  & \celladderror{44.19}{38}{86}
  & \celladderror{40.70}{35}{86}
  & \celladderror{42.11}{32}{76}
  & \celladderror{19.35}{12}{62}
  & \celladderror{38.33}{23}{60}
  & \celladderror{31.33}{26}{83}
  & \celladderror{55.41}{41}{74} \\

DeepSeek-R1-Distill-Qwen-7B
  & \celladderror{44.45}{10}{22}
  & \celladderror{86.05}{37}{43}
  & \celladderror{77.27}{34}{44}
  & \celladderror{76.19}{32}{42}
  & \celladderror{37.38}{40}{107}
  & \celladderror{60.71}{68}{112}
  & \celladderror{60.36}{67}{111}
  & \celladderror{63.55}{68}{107}
  & \celladderror{29.89}{26}{87}
  & \celladderror{46.81}{44}{94}
  & \celladderror{44.57}{41}{92}
  & \celladderror{77.08}{74}{96} \\

DeepSeek-R1-Distill-Qwen-14B
  & \celladderror{62.96}{17}{27}
  & \celladderror{79.07}{34}{43}
  & \celladderror{81.82}{36}{44}
  & \celladderror{82.22}{37}{45}
  & \celladderror{41.44}{46}{111}
  & \celladderror{59.83}{70}{117}
  & \celladderror{62.71}{74}{118}
  & \celladderror{70.94}{83}{117}
  & \celladderror{31.18}{29}{93}
  & \celladderror{43.75}{42}{96}
  & \celladderror{44.21}{42}{95}
  & \celladderror{72.92}{70}{96} \\

DeepSeek-R1-Distill-Qwen-32B
  & \celladderror{67.74}{21}{31}
  & \celladderror{87.80}{36}{41}
  & \celladderror{84.44}{38}{45}
  & \celladderror{88.37}{38}{43}
  & \celladderror{54.05}{60}{111}
  & \celladderror{63.48}{73}{115}
  & \celladderror{68.07}{81}{119}
  & \celladderror{71.07}{86}{121}
  & \celladderror{30.43}{28}{92}
  & \celladderror{47.42}{46}{97}
  & \celladderror{47.41}{44}{93}
  & \celladderror{78.89}{71}{90} \\
  \hline

GPT-4o mini
  & \celladderror{62.50}{25}{40}
  & \celladderror{88.89}{40}{45}
  & \celladderror{88.89}{40}{45}
  & \celladderror{88.89}{40}{45}
  & \celladderror{77.69}{94}{121}
  & \celladderror{70.16}{87}{124}
  & \celladderror{71.54}{88}{123}
  & \celladderror{76.80}{96}{125}
  & \celladderror{45.33}{34}{75}
  & \celladderror{70.83}{68}{96}
  & \celladderror{77.08}{74}{96} 
  & \celladderror{81.44}{79}{97} \\
  \bottomrule
\end{tabular}

\smallskip
\raggedright
\footnotesize
\textbf{Legend:} UD = Update Description; UD+Doc = Update Description + API Documentation; +CoT = UD+Doc+CoT; +CoT/SR = UD+Doc+CoT+SR.
\end{table*}


\textbf{When given only an update description (UD), LLMs often fail to generate usable code examples. On average, 25.36\% of outputs completely ignore the provided API update specification. Among the cases that adopt the updated API, only 42.55\% are executable.} Table~\ref{tab:results_adoption_rate} and ~\ref{tab:ExecRate_results} present the adoption rate and executable rate of each studied model across the three patterns, respectively. The UD-only setting exposes severe limitations across both metrics. On the adoption rate, several models fall below 50\% in at least one pattern under UD alone,  which indicates these models generate code that completely ignores the provided update specification less than half the time when given only a textual description of the change. For instance, CodeLlama-13B (15.56\%), DeepSeek-R1-1.5B (44.44\%), and 7B (48.89\%)suffer from a low adoption rate in P1. 
The executable rate under UD is even more alarming.  24 out of 33 model-pattern combinations fall below 50\%, such as CodeLlama-7B on P1 (11.11\%), DeepSeek-R1-1.5B on P3 (19.35\%), and DeepSeek-R1-1.5B on P2 (22.37\%). This is expected: without documentation, models have no information about a newly introduced API and must hallucinate its interface entirely, almost always incorrectly. 

\textbf{Providing API documentation substantially improves models’ ability to adopt updated APIs (adoption rate) from 74.64\% to 92.87\% (an improvement of 24.42\%) on average, and improves the executable rate from 42.55\% to 66.36\% with an improvement of 55.96\%.} The introduction of API documentation produces the largest single improvement in the entire prompting pipeline for both metrics. Adoption rate gains are most dramatic in P1 (60.07\% improvement), where documentation directly supplies the replacement API that models were failing to identify from the description alone. Executable rate gains are largest in P1 (76.52\% improvement) and P3 (65.44\% improvement), the two patterns where models have the least prior knowledge to rely on. 
While both metrics improve substantially, the executable rate remains below 70\% even with documentation. This establishes that documentation is necessary but not sufficient for reliable code generation under API evolution.

\textbf{Different API update patterns pose different levels of difficulty for LLMs, with API modification (P2) being the most challenging.}
Across all models, P2 consistently shows the lowest adoption rate (87.14\% under UD+Doc) and the lowest executable rate (58.19\%). In contrast, P1 (API deprecation) achieves higher performance, as it mainly requires replacing deprecated APIs with existing alternatives—many of which are already present in the models’ parametric knowledge (e.g., 27/45 the deprecated APIs in our dataset predate Dec 2023). P3 (API addition) performs moderately, as it relies on learning entirely new APIs from external context. 
P2 is the hardest, possibly because it requires modifying existing usage patterns (e.g., parameter changes or argument reordering) while keeping the same interface. These subtle changes create stronger conflicts between external context and prior knowledge, causing models to retain outdated parameter usage. This actually is evidenced in our failure analysis in Section~\ref{sec:rq4}.


\begin{figure}[t]
\centering
\begin{minipage}{\columnwidth}
    \centering
    \includegraphics[width=\linewidth]{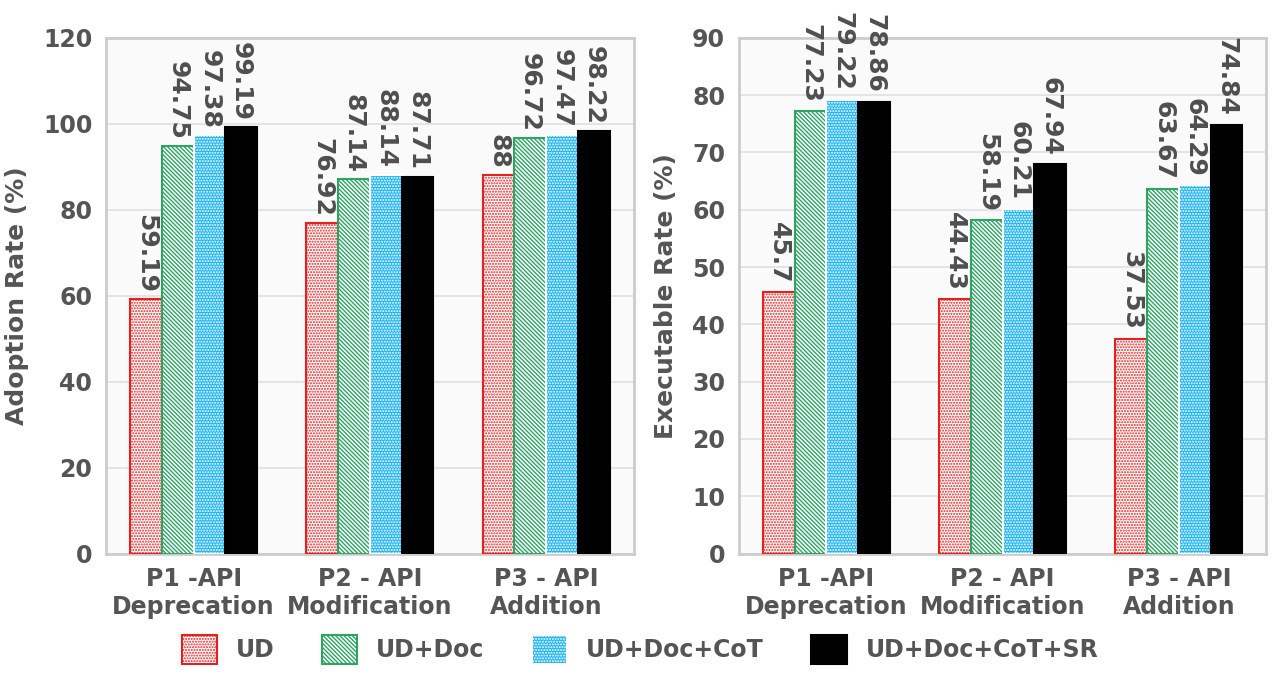}
\end{minipage}
\caption{The average adoption rate (left) and executable rate (right) across all models on each pattern when providing different context (UD and Doc) and using different prompting techniques (CoT and SR).}\label{fig:across_patterns}
\end{figure}

\rqboxc{\textbf{RQ summary:} LLMs struggle to follow externally provided API update specifications without full documentation, with only 74.64\% adoption and 42.55\% executable rate under update description alone. Adding API documentation yields the largest improvement, raising adoption to 92.87\% and executable rate to 66.36\%. API modification (P2) remains the most challenging pattern, even though API documentation is provided across both metrics (87.14\% adoption rate and 58.19\% adoption rate).}

\subsection{RQ2 - Impact of model family and  model size}\label{sec:rq2}


\textbf{We observe different trends in how effectively models follow API update specifications across different models. GPT-4o-mini achieves the highest adoption rate and executable rate.} As Table~\ref{tab:results_adoption_rate} presented, GPT-4o mini is the strongest single model overall, reaching 98.61\% adoption rate and 76.63\% executable rate under UD+Doc across all model families. Among open-source families, all models achieve nearly similar adoption rates under UD+Doc, DeekSeek-R1 91.39\%, CodeLmama 94.06\%, and DeekSeek-coder 91.74\%. However, on executable rate, DeepSeek-coder leads at 75.76\%, while CodeLlama (61.58\%) and DeepSeek-R1 (60.34\%) are substantially lower. This divergence between adoption rate and executability indicates that they follow the updated API specification, but their generated code is more prone to execution failures, particularly on P2 (API Modification) for CodeLlama-13B (14.29\%) and on P3 (API Addition) for all R1 models (38.33 – 47.42\%).

\textbf{Scaling parameter size helps improve adoption rate within families, but does not resolve family-specific weaknesses on executability.} Under UD+Doc, four out of nine pattern-model family combinations produce a monotone improvement of adoption rate as the model parameter size increases. Typically, the smallest models usually performed the worst among all families (7 out of the 9 pattern-model family combinations). For instance, within Deepseek-coder, adoption rate under UD+Doc scales reasonably well from 1.3B (71.9\%) to 33B (90.6\%) in P2. Regarding executable rate, monotonic improvements are observed only for the DeepSeek-R1 model family, increasing from 44.19\% at 1.5B parameters to 63.48\% at 32B in P2.

\rqboxc{\textbf{RQ summary:}  We observe different trends in how effectively models follow
API update specifications across different models. Scaling parameter size generally improves adoption, but does not consistently resolve family-specific weaknesses in executability.}

\subsection{RQ3 - Improvement by reasoning-based prompting engineering}\label{sec:rq3}


\noindent\textbf{CoT and SR yield marginal gains in adoption rate but substantial improvements in executable rate (11.33\%), with self-reflection (SR) contributing more than chain-of-thought (CoT) to executability.}   Figure~\ref{fig:across_patterns} presents the average adoption rate and executable rate across all patterns after applying CoT and SR. Overall, among the 33 model-pattern combinations, applying CoT/SR improves the adoption rate in 20 combinations and the executable rate in 26 combinations, respectively. CoT provides only marginal improvement over UD+Doc in both metrics, i.e., 1.57\% gain in adoption rate and 2.34\% in executable rate, indicating that explicit reasoning steps alone offer limited benefit once full API documentation is already supplied. SR, applied on top of CoT, delivers negligible further gains in adoption rate (0.77\%), but produces a substantially larger improvement in executable rate (9.00\% improvement over +CoT, or 11.33\% improvement over UD+Doc). This asymmetry suggests that SR's self-verification loop is particularly effective at catching implementation-level errors, rather than high-level failures to follow the API update specification, which are already largely resolved by the documentation itself. In other words, SR helps models not just adopt updates, but use them correctly enough to produce runnable code.



\textbf{The benefit of CoT and SR is highly pattern-dependent. SR is particularly helpful in improving the executable rate in P2 and P3.} For adoption rate, gains are concentrated in P1 (API Deprecation). CoT+SR improves P1's adoption rate from 94.75\% to 99.19\%, while both contribute almost nothing to P2 and P3.  For executable rate, the trend is the opposite: P2 and P3 are where Cot+SR makes its biggest impact. SR+CoT gains 16.8\% from UD+DOC on P2 and 17.5\% on P3, while P1 only improves by 2.1\%. As observed from Figure~\ref{fig:across_patterns}, the majority of the benefit is contributed by SR. One explanation is that SR converts a generation task into a verification task. The initial generation under UD+Doc is prone to structural errors driven either by knowledge anchoring or by hallucination, which is evidenced in Section~\ref{sec:rq4}. SR's self-check creates a second opportunity for the model to catch structural code errors before finalizing the output. Errors that are invisible during forward generation but detectable during backward verification against the provided specification. 


\rqboxc{\textbf{RQ summary:} Reasoning-based prompting offers limited gains in adoption rate (+1.57\% for CoT, +0.77\% for SR) but substantially improves executable rate (+11.33\% overall), with SR contributing the most. Benefits are pattern-dependent: SR is most effective for P2 and P3 executability (+16.8\% and +17.5\%).}

\subsection{RQ4 - Failure case analysis}\label{sec:rq4}

\begin{figure}[t]
\centering
\begin{minipage}{0.99\columnwidth}
    \centering
    \includegraphics[width=\linewidth]{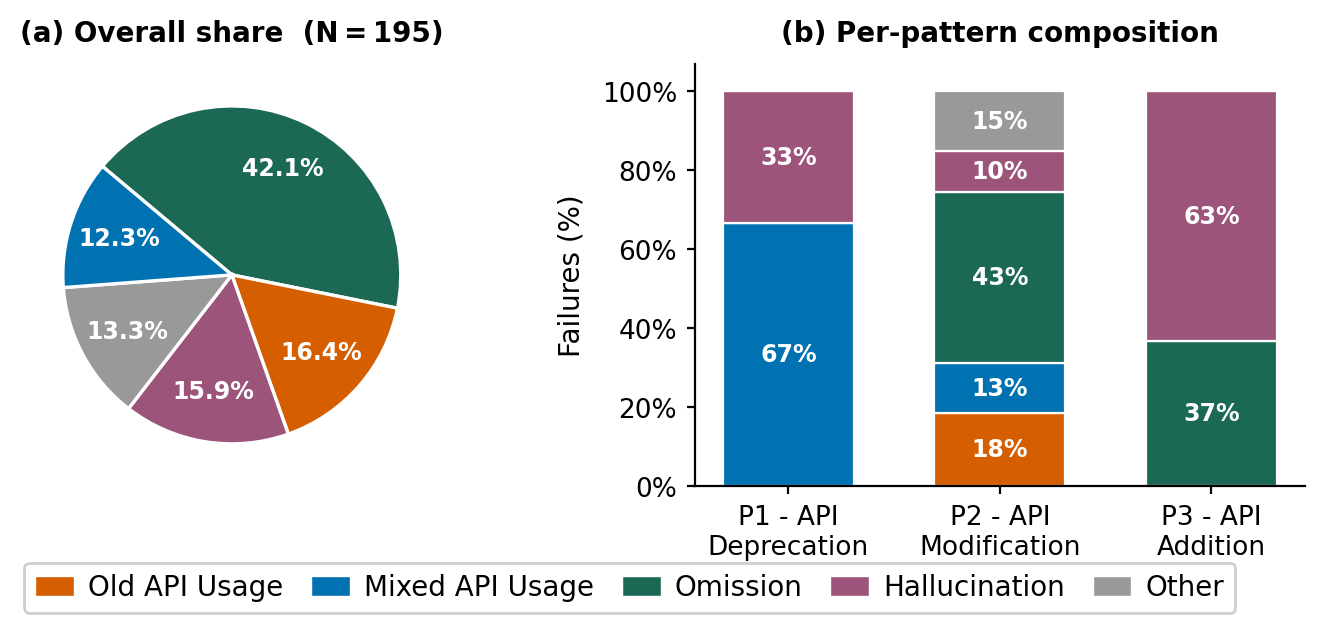}
\end{minipage}
\caption{Distribution of failure types in LLM-generated code where the API update is not adopted at all: (a) aggregated results, and (b) broken down by API update pattern.} 
\vspace{-0.1in}\label{fig:failure_when_not_followed}
\end{figure}

\noindent\textbf{Failure cases where the API update is not adopted at all are primarily driven by three error types: complete omission of the provided update specification (42.1\%), use of deprecated or outdated APIs (16.4\%), and mixed usage of both new and deprecated APIs (12.3\%).}
Figure~\ref{fig:failure_when_not_followed} presents the distribution of failure types across the three update patterns. The most prevalent failure type is \textit{Omission}, occurring in 82 out of 195 cases (42.1\%), where the model entirely disregards the updated API specification provided in the prompt and generates code as if no external context were given. Figure~\ref{fig:api-hallucination} illustrates a representative example: when prompted to generate code using the newly introduced \texttt{numpy.unique\_values} API from NumPy 2.0.0 (June 2024), GPT-4o mini ignored the provided API documentation entirely — even under CoT and SR prompting — and instead invoked an older API with a similar name. This suggests that when a newly introduced API closely resembles an existing one in name or functionality, models tend to anchor on their parametric knowledge rather than follow the provided specification. \textit{Old API Usage} and \textit{Mixed API Usage} account for 16.4\% and 12.3\% of failures, respectively, indicating that in a non-trivial portion of cases, models either revert entirely to outdated usage or produce hybrid code that interleaves deprecated and updated API calls.
The distribution of failure types varies considerably across update patterns. For P1 (API Deprecation), the adoption rate is highest among all three patterns after applying CoT+SR, with only 3 failure cases remaining. Two of these involve mixed usage where both the deprecated and recommended replacement APIs appear in the same code example, suggesting partial but incomplete context following. P2 (API Modification) is the most challenging pattern overall, with all five failure categories present. The dominant failure type is parameter-level omission — where the model adopts the correct function name but ignores modifications to its parameters — accounting for 43.4\% of P2 failures. This suggests that models are more responsive to function-level changes than to finer-grained parameter-level updates, even when the latter are explicitly documented in the provided context. P3 (API Addition) exhibits a distinct failure profile: in 63\% of failure cases, models hallucinate a non-existent API — fabricating a plausible but invalid function rather than using the newly introduced one specified in the documentation — while the remaining 37\% constitute pure omission cases where the provided API is ignored entirely. The high hallucination rate in P3 suggests that when models encounter an unfamiliar API name in the prompt, they are prone to substituting a confabulated alternative rather than faithfully adopting the documented one.

\definecolor{docpurple}{HTML}{83b11b}
\definecolor{groundgreen}{HTML}{3E883B}
\definecolor{llmred}{HTML}{B44A2F}
\definecolor{codebg}{RGB}{248,248,248}
\definecolor{kw}{HTML}{008000}
\definecolor{bi}{HTML}{0055AA}
\definecolor{hlred}{HTML}{FDDDDD}

\lstdefinestyle{pycol}{
  language=Python,
  basicstyle=\ttfamily\scriptsize,
  keywordstyle=\color{kw}\bfseries,
  stringstyle=\color{purple},
  commentstyle=\color{gray}\itshape,
  showstringspaces=false,
  breaklines=true,
  columns=flexible,
  escapeinside={(*@}{@*)},
  xleftmargin=3pt,
  xrightmargin=2pt,
  aboveskip=1pt,
  belowskip=1pt,
  frame=none,
  backgroundcolor={},
}

%

\definecolor{paperRed}{RGB}{153, 0, 0}
\definecolor{paperGray}{RGB}{245, 245, 245}
\definecolor{codeFrame}{RGB}{200, 200, 200}

\tcbset{
  academicbox/.style={
    enhanced,
    width=\columnwidth,
    boxrule=0.6pt,
    colframe=codeFrame,
    colback=white,
    arc=0mm, 
    fonttitle=\sffamily\bfseries\small,
    coltitle=black,
    colbacktitle=paperGray,
    attach boxed title to top left={yshift=-2mm, xshift=2mm},
    boxed title style={colframe=codeFrame, colback=paperGray, sharp corners},
    top=0mm,    
    bottom=0mm, 
  }
}

\begin{figure}[t]
\begin{tcolorbox}[academicbox, title={Failure case- Omission}]
\begin{lstlisting}[
    language=Python,
    basicstyle=\ttfamily\scriptsize,
    keywordstyle=\color{blue},
    commentstyle=\color{gray},
    breaklines=true,
    escapeinside={(*@}{@*)}
]
import numpy as np
 
# Create an array with some duplicate values
array = np.array(
    [1, 2, 2, 3, 4, 4, 5, 5, 5])
 
# Use numpy.unique to get unique values
(*@\colorbox{hlred}{unique\_values = np.unique(array)}@*)
 
# Print the unique values
print("Unique values in the array:",
      unique_values)
\end{lstlisting}


\begin{tabular}{rp{0.8\columnwidth}}
    \textcolor{paperRed}{\ding{55}} & \small Ignored the new API: \colorbox{hlred}{\texttt{\scriptsize np.unique\_values}}.
\end{tabular}
\end{tcolorbox}
\vspace{-0.2in}
\caption{Example of omitting an API update (i.e., introducing a new API \colorbox{hlred}{\texttt{numpy.unique\_values}}) in LLM-generated code from GPT4o-mini, where it ignored the specified API and used the API name as a variable.}
\vspace{-0.1in}
\label{fig:api-hallucination}
\end{figure}

\begin{figure}[t]
\centering
\begin{minipage}{0.99\columnwidth}
    \centering
\includegraphics[width=\linewidth]{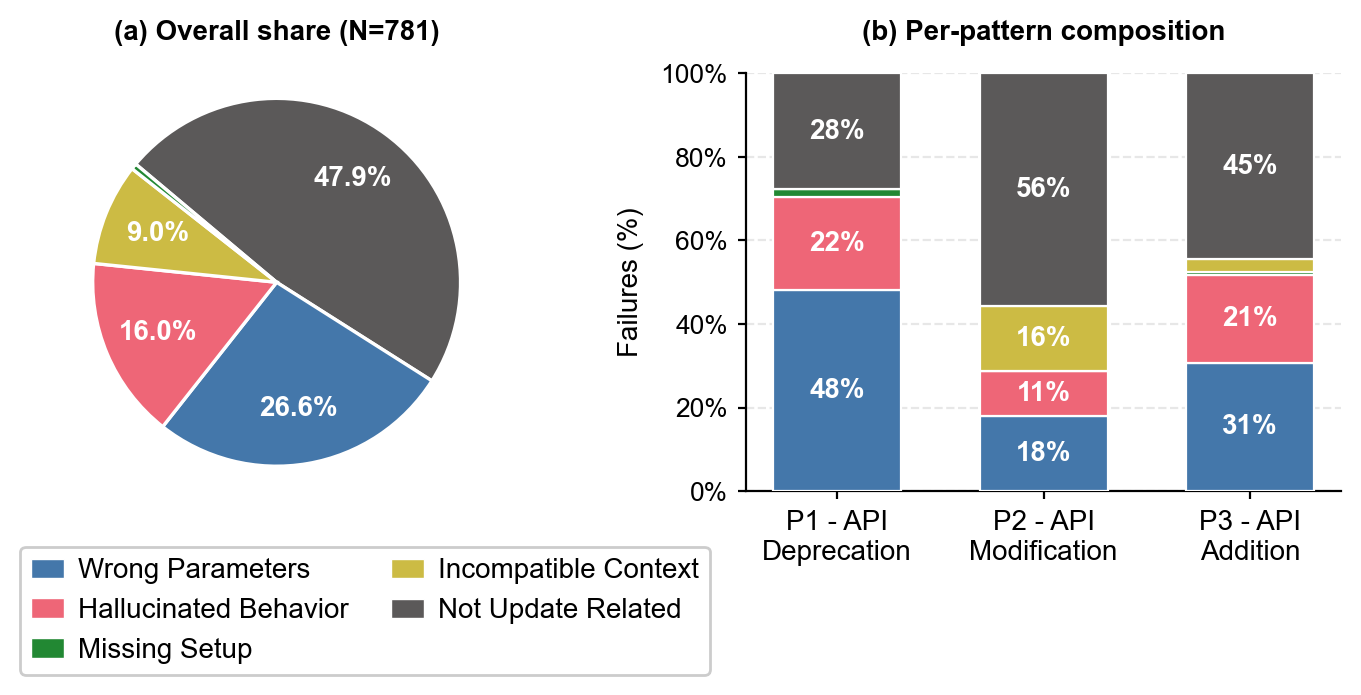
}
\end{minipage}
\caption{Distribution of failure types in LLM-generated code that adopted the API update but fails to execute: (a) aggregated results, and (b) broken down by API update pattern.}\label{fig:failure_when_followed}
\end{figure}

\noindent\textbf{Even when LLMs adopt the correct updated API, over half of execution failures stem from incorrect API usage rather than unrelated implementation errors, with wrong parameter usage and hallucinated API behavior being the dominant causes.}
In cases where the model successfully adopted the updated API at the selection level, a non-trivial portion of generated code still failed at execution time, revealing that API adoption alone does not guarantee correct usage. Figure~\ref{fig:failure_when_followed} presents the distribution of execution failure causes under our best-performing configuration (UD+Doc+CoT+SR). Failures fall into two broad categories: those attributable to the API update itself, and those unrelated to it.
The majority of execution failures (52.1\%) are directly related to incorrect API usage. The most prevalent update-related failure type is \textit{Wrong Parameters} (26.6\%), which encompasses both the use of unsupported parameters and the omission of required ones. This suggests that while models can often learn the correct API to invoke from the provided context, they struggle to faithfully reproduce its full parameter specification — particularly when the updated signature differs substantially from what the model has internalized from pre-training. A further notable failure type is \textit{Hallucinated Behavior} (the third most prevalent category), where the model invokes the correct API but makes incorrect assumptions about its runtime behavior — for example, assuming an incorrect return type or chaining a method that does not exist on the returned object. These failures are especially insidious as they are unlikely to be caught through static inspection and only surface at execution time. Figure~\ref{fig:hallucination_v2} illustrates a representative instance of a hallucinated parameter value error in which GPT-4o-mini correctly identified the target API and its valid set of parameters. However, the value assigned to the rule parameter — '\textit{genz-malek}', does not exist in the API's defined set of accepted rule identifiers. 
The remaining 47.9\% of execution failures fall into the \textit{Not Update Related} category, primarily attributable to syntax errors and other implementation issues independent of the API update specification. 

\begin{figure}[t]
\begin{tcolorbox}[academicbox, title={Failure case- Hallucination}]
\begin{lstlisting}[
    language=Python,
    basicstyle=\ttfamily\scriptsize,
    keywordstyle=\color{blue},
    commentstyle=\color{gray},
    breaklines=true,
    escapeinside={(*@}{@*)}
]
from scipy.integrate import cubature 

def f(x, n):
    return x[:, np.newaxis] ** n[np.newaxis, :]

res = cubature(
    f, a=[0], b=[1],
    args=(np.arange(10),),
    (*@\colorbox{red!20}{rule='genz-malek'}@*),
    rtol=1e-6, atol=1e-6)
print("Estimate:", res[0])
\end{lstlisting}


\begin{tabular}{rp{0.8\columnwidth}}
    \textcolor{paperRed}{\ding{55}} & \small Added a non-existing rule- \texttt{'genz-malek'}.
\end{tabular}
\end{tcolorbox}
\caption{Example of hallucination: the LLM invented a non-existent rule \colorbox{hlred}{\texttt{'genz-malek'}} for scipy.integrate.cubature.}
\label{fig:hallucination_v2}
\end{figure}

\rqboxc{\textbf{RQ summary:} LLM failures fall into two levels. At the adoption level, omission (42.1\%), old API usage (16.4\%), and hallucinated APIs (63\% of P3 failures) dominate. At the execution level, over half of failures stem from incorrect API usage, such as parameters (26.6\%) and hallucinated behavior (16\%).}

\section{Discussion}\label{sec:dis}

\subsection{Implications of our findings}\label{sec:implications}

\noindent\textbf{API documentation should be treated as a first-class input in LLM-assisted development workflows.}
Our findings consistently show that the single largest performance driver across all models, patterns, and metrics is the provision of structured API documentation — not model size, model family, or reasoning-based prompting. Adding documentation to the prompt raises adoption rate by 24.42\% and executable rate by 57.42\% relative to update descriptions alone. This has a direct implication for developer tooling: IDE plugins, copilot integrations, and RAG-based coding assistants should be designed to automatically retrieve and inject up-to-date API documentation into the prompt context whenever a code generation task involves a library dependency. Relying on developers to manually supply this context, or assuming that a model's parametric knowledge is sufficient, leads to systematically degraded output quality under API evolution.

\noindent\textbf{Self-reflection prompting should be prioritized for tasks involving API modification and newly introduced APIs.}
While reasoning-based prompting offers only marginal gains in adoption rate (2.34\% for CoT/SR overall), it delivers substantial improvements in executable rate — particularly for P2 (API Modification, 16.8\%) and P3 (API Addition, 17.5\%) relative to UD+Doc alone. This suggests that self-reflection is most valuable precisely in the cases where parametric knowledge is most likely to interfere: when an existing API has been subtly changed, or when a newly introduced API is unfamiliar to the model. Tool builders and prompt engineers should therefore apply SR selectively — prioritizing it for modification and addition patterns rather than applying it uniformly — and future work should explore lightweight triggers that detect when SR is likely to yield the greatest benefit based on the type and recency of the API change.

\noindent\textbf{The community needs continuously evolving benchmarks that track API changes beyond model training cutoffs to evaluate LLMs explicitly account for post-training knowledge gaps.}
Our results show that LLMs fail in consistent, predictable ways when asked to use APIs introduced after their training cutoff. 42.1\% of non-adoption failures involve the model ignoring the provided specification entirely or using old APIs (16.4\%). 42.6\% of execution failures stem from wrong parameter usage or hallucinated behavior. These are not random errors. Instead, they follow directly from models defaulting to memorized API patterns even when the prompt explicitly provides updated information. Yet existing code generation benchmarks such as HumanEval and SWE-bench do not test this scenario, as their tasks are drawn from within the model's training distribution. We therefore call for benchmarks that are continuously updated alongside real library releases, so that the community can track how models and prompting strategies hold up as APIs evolve over time. On the model training side, incorporating API migration examples — particularly for post-cutoff library versions — into instruction tuning datasets could help reduce models' tendency to anchor on stale usage patterns when more current information is available in context.

\subsection{Threats to validity}
\noindent\textbf{Threats to External Validity.} Threats to external validity concern the generalizability of our findings beyond the scope of our study. First, our benchmark is constructed from eight Python libraries, which are drawn from the data science and machine learning domains. While these libraries are widely used and represent a diverse range of API evolution patterns, our findings may not fully generalize to other domains and programming languages, where API change characteristics and documentation quality may differ substantially. Second, we evaluate eleven LLMs spanning three open-source families and one proprietary model. While this covers a representative range of model sizes and architectures, other model families and more recent models released after our study period may exhibit different behaviors under API evolution. We encourage future studies to replicate our methodology across additional languages, domains, and model families to broaden the generalizability of these findings.

\noindent\textbf{Threats to Internal Validity.}
Threats to internal validity concern potential confounds that may affect the accuracy and reliability of our measurements. First, our dataset construction relies on LLM-based extraction of API changes from release notes, which may introduce extraction errors or misclassifications into the three update patterns. To mitigate this, two authors independently inspected all extracted instances and resolved disagreements through discussion, and instances with incomplete or ambiguous documentation were discarded. Second, we adopt an LLM-as-a-Judge approach to automatically assess API Adoption Rate and conduct failure type analysis at scale, using GPT-5 mini as the judge model. This may introduce noise and bias our findings. To mitigate this, we manually annotated sampled instances, and achieved 91.4\% and 90\% accuracy, indicating this method is reliable.

\section{Conclusion}\label{sec:conclusion}

This paper presents a systematic study of LLM code generation under the condition of context-memory conflict caused by API evolution. Our results reveal that while providing external documentation significantly improves API adoption, models still struggle to override stale parametric knowledge, often resulting in low execution success. We demonstrate that reasoning-based prompting, specifically Self-Reflection, is an effective mechanism for improving model performance when conflicts occur. Our failure analysis highlights that adoption failures are frequently due to the omission of updated API or sticking to old APIs. At the execution level, over half of failures stem from incorrect API usage, such as parameters and hallucinated behavior. Our findings highlight the persistence of outdated patterns from LLMs, even API update specification is provided, and emphasize the need for evolution-aware benchmarks and techniques.


\section{Data Availability Statement}
We make our code and data public in the repo~\cite{conflict_codegen_unannonymous}.

\bibliographystyle{ACM-Reference-Format}

\bibliography{main.bbl}



\end{document}